# Evidence for a strain tuned topological phase transition in ZrTe$_5$

Joshua Mutch[1], Wei-Chih Chen[2], Preston Went[1], Tiema Qian[1], Ilham Zaky Wilson[1], Anton Andreev[1], Cheng-Chien Chen[2*], Jiun-Haw Chu[1*]

[1]Department of Physics, University of Washington, Seattle, Washington 98195, USA

[2]Department of Physics, University of Alabama at Birmingham, Birmingham, Alabama 35294, USA

*Correspondence to: jhchu@uw.edu (JHC), chencc@uab.edu (CCC)

**Abstract:** A phase transition between topologically distinct insulating phases involves closing and reopening of the bandgap. Close to this topological phase transition, the bulk energy spectrum is characterized by a massive Dirac dispersion, where the mass plays the role of bandgap. Here we report the observation of a non-monotonic strain dependence of resistivity and negative longitudinal magnetoresistance in ZrTe$_5$, which is known to host massive Dirac Fermions in the bulk. This non-monotonic strain dependence is consistent with the closing and reopening of the bandgap at the Brillouin-zone center, indicative of a topological phase transition. This observation suggests that the topological state of ZrTe$_5$ is highly sensitive to uniaxial stress. Our study presents a promising platform for continuous *in-situ* control of nontrivial topological properties of materials.

**One Sentence Summary:** Uniaxial stress is used to close and reopen the bandgap in ZrTe$_5$, indicative of a strain-tuned topological phase transition.

**Main Text:** Appreciation of the topological aspects of band structure has fundamentally changed the way we understand the electronic structure of solids. Band insulators with time reversal symmetry can be classified into normal insulator (NI), weak topological insulator (WTI) and strong topological insulator (STI) based on their Z$_2$ topological indices (*1-3*). Changing Z$_2$ topological indices requires closing and reopening the bandgap. Therefore, topologically distinct insulating phases are separated by a gapless state. If inversion symmetry is broken, then the gapless state is a Weyl semimetal that is robust against small perturbations. In the presence of inversion symmetry, a gapless Dirac semimetal only exists at the phase boundary. The relationship between these phases is summarized in the general phase diagram proposed by Murakami et al.(*2, 3*), as shown in Fig. 1A.

Many of these topological phases have been intensively studied in the past decade (*4-9*). In comparison, the transitions between these phases are less explored. It has been demonstrated that topological phase transitions can be induced either by chemical doping (*10, 11*) or thermal lattice expansion (*12, 13*). Nevertheless, the precise *in-situ* control of topological indices remains an outstanding challenge, which is an important first step towards building functional devices based on non-trivial topological properties. In this work, we first performed *ab initio* band structure calculations to show that it is possible to induce a topological phase transition in ZrTe$_5$ by applying less than a percent of anisotropic strain. We then experimentally studied the transport properties of single crystalline ZrTe$_5$ as a function of *in-situ* tunable anisotropic strain. We observed non-monotonic strain dependence of resistivity and negative longitudinal magnetoresistance, which provides strong evidence for the closing and reopening of the bandgap, and hence a STI to WTI transition.

ZrTe$_5$ is a van der Waals (vdW) layered material crystallized in the Cmcm orthorhombic space group. Each layer consists of ZrTe$_3$ chains extending along the *a*-lattice direction, and the layers are stacked along the *b*-lattice direction (Fig. 1B). The material received significant interest because its mono-layer form was predicted theoretically to be a large bandgap quantum spin Hall insulator (*14*). It was also suggested that the three-dimensional bulk band structure is very close to the phase boundary between WTI and STI (*14-16*). Early experimental studies include optical conductivity (*17*), Berry phase of quantum oscillations (*18, 19*), photoemission and negative longitudinal magnetoresistance associated with the chiral anomaly (*20*). These works are consistent with a Dirac semimetal-like band structure in the bulk, with a single Dirac point at the center of the Brillouin zone. Unlike topological Dirac semimetals such as Na$_3$Bi or Cd$_3$As$_2$, there is no additional crystalline symmetry to protect the Dirac point. The Dirac dispersion in ZrTe$_5$ is formed because of its proximity to WTI-STI phase boundary. Indeed, more recent spectroscopy measurements revealed that the band structure is better described by a massive Dirac dispersion, where mass plays the role of bandgap. The bandgap size measured by different experiments varies, ranging from 10 to 80 meV, and there are conflicting reports on whether the material is a WTI or STI (*21-24*). Despite the disagreement, the small size of the bandgap suggests that it is a promising candidate for dynamically tunable band inversion with external controls, such as strain.

To explore this possibility, we first used density functional theory (DFT) to calculate the band structure of ZrTe$_5$. Fig. 1C shows the size of the bandgap $E_g$ at the Γ point as functions of strain (%) along the *a* and *c* lattice directions, using the previously reported experimental lattice constants (*a* = 3.97976 Å , *c* = 13.6762 Å ) as the zero strain values (*25*). The $E_g$ contour shows a V-shaped valley, with the minimum of the valley (corresponding to $E_g = 0$) extending along the diagonal direction. Z$_2$ topological indices have also been computed for each strain state, and the $E_g = 0$ line is indeed the phase boundary between STI and WTI (see Fig. S7). Fig. 1C reveals a highly anisotropic strain dependence of $E_g$: the steepest gradient of $E_g$ is along the direction where $\epsilon_{aa}$ and $\epsilon_{cc}$ have opposite sign, which can be induced by a uniaxial stress along the *a*-lattice direction. The Poisson's ratio $\epsilon_{cc}/\epsilon_{aa} = -0.25$ indicated as the grey arrow in Fig. 1C is obtained by fully relaxed vdW-DFT calculation (*26*). With this Poisson's ratio, it requires less than a percent of $\epsilon_{aa}$ to reach the WTI-STI phase boundary. We note that there is uncertainty in the DFT bandgap size for a given set of lattice constants. For example, several spectroscopy measurements reported $E_g$ as low as 10meV, which is significantly lower than the 60 meV DFT bandgap in the zero strain state (*22, 23*). In other words, the actual strain to achieve band inversion could be even smaller (see Fig. 1D).

We then experimentally investigated the strain dependence of transport properties of ZrTe$_5$ single crystals. The electrical transport of ZrTe$_5$ is dominated by the conduction of bulk Dirac fermions, so it is difficult to use the surface state transport as a probe of topological phase transition. Nevertheless, as the bulk energy bandgap closes and reopens, the mass of Dirac fermions is also modulated accordingly, leading to a non-monotonic strain dependence of resistivity. The effect of bandgap opening on resistivity is stronger if the chemical potential is closer to the Dirac point. Therefore, we used the flux method to grow single crystals of ZrTe$_5$ (*27*). The flux method is known to yield crystals closer to perfect chemical stoichiometry and with much lower carrier density (*p*-type $10^{15}$ cm$^{-3}$) compared to vapor transport grown crystals (*n*-type $10^{17}$ cm$^{-3}$) (*28*). Fig. 2A shows resistivity as a function of temperature for free standing ZrTe$_5$ single crystals before mounting on the strain apparatus. The insulating temperature dependence is consistent with other flux grown crystals in the literature, suggesting that at base temperature the chemical potential of our samples lies just slightly below the valence band maximum (*24, 29*).

Uniaxial stress was applied along the *a*-axis of ZrTe$_5$ using a piezoelectric apparatus introduced by Hicks et al. (*30, 31*), as shown in Fig. 2B. A long needle-like crystal was glued across a gap in the apparatus onto titanium pieces. Expanding or contracting the piezostacks changes the gap size (*L*) and applies uniaxial stress to the crystal. The uniaxial stress induces strain along all three crystallographic directions. The strain along the *a*-axis $\epsilon_{aa}$ is equal to $\alpha \Delta L/L$, where the change of the gap size $\Delta L/L$ is estimated by a strain gauge glued on the piezostacks. The constant $\alpha$ takes into account a strain relaxation effect, which is estimated by finite element analysis and is typically about 0.8 (*27*). The strain along the *b*-axis and *c*-axis are determined by the Poisson's ratio. Resistivity of the sample was measured along the *a*-axis via a conventional 4-point matchstick geometry.

Fig. 2C shows the resistivity as a function of $\epsilon_{aa}$ measured at *T = 2K*. The strain dependence of resistivity is non-monotonic. For all samples measured, the resistivity showed a minimum at a critical strain $\epsilon_{min}$, with a highly symmetric quadratic dependence close to $\epsilon_{min}$. There is an uncertainty in determining the zero-strain state, i.e. the absolute value of $\epsilon_{min}$, due to an uncontrolled strain generated by the mismatch of thermal contraction between the sample and apparatus. Nevertheless, based on a detailed analysis we estimated $\epsilon_{min} < 0.1\%$ (*27*). For all the data shown here, $\epsilon_{aa}$ is measured from $\epsilon_{min}$. We note that although the size of non-linear response, i.e. the quadratic coefficient, varies from sample to sample, the appearance of the resistivity minimum is a robust phenomenon. It is well known that the resistivity of semiconductors can have a large *linear* response to strain due to its sensitivity to the position of band edges as a function of strain. However, such a *non-monotonic* strain dependence is rather unusual, but it can be naturally explained by a strain-induced bandgap-closing and reopening. To investigate this possibility, the strain dependence of longitudinal magnetoresistance (*I* ∥ *B* ∥ ***a***) was also measured.

A negative longitudinal magnetoresistance has been observed in ZrTe$_5$ and attributed to the chiral anomaly (*20*). This effect was initially proposed for gapless Weyl semimetals (*32, 33*), yet for a gapped Dirac semimetal essentially the same mechanism could still apply in the semiclassical regime provided that $E_g/E_F \ll 1$ (*34*). In a gapped Dirac semimetal electron helicity plays the same role as chirality in Weyl semimetals. The helicity relaxation rate increases due to bandgap opening. In other words, we expect a suppression of negative longitudinal magnetoresistance (or positive longitudinal magnetoconductivity) when the bandgap reopens. Fig. 2D shows the longitudinal magnetoconductivity $\Delta\sigma = \sigma(B) - \sigma(B = 0)$ for different strain states. We focus on the low field semiclassical regime (*B<1T*), since for *B* ∥ ***a*** at *1T* the low carrier density sample ($10^{15}$ cm$^{-3}$) is approaching the estimated quantum limit (*18, 28*). Within this field range, the magnetoconductance is positive except showing a small dip near zero field (*35*). At $\epsilon_{aa} = \epsilon_{min}$ the magnetoconductance reaches its maximum, and it is progressively suppressed when sample was strained away from $\epsilon_{aa} = \epsilon_{min}$. This non-monotonic strain dependence of longitudinal magnetoresistance provides another strong evidence for the vanishing of the bandgap at $\epsilon_{aa} = \epsilon_{min}$.

Below we provide a quantitative description of the strain dependence of resistivity, i.e. elastoresistivity. Strain affects the parameters of the low-energy electron Hamiltonian. In ZrTe$_5$ a $k \cdot p$ Hamiltonian was constructed based on the symmetry constraints, and it was applied successfully to the fitting of magneto-optics spectrum (*22, 23, 36*). The $k \cdot p$ Hamiltonian gives rise to a massive Dirac dispersion:

$$E(\boldsymbol{k}) = \pm \sqrt{m^2 + \sum_{\alpha=x,y,z} \hbar^2 v_\alpha^2 k_\alpha^2}$$

where $m$ is half of the bandgap $E_g = 2|m|$ and $v_\alpha$ are the Fermi velocities. Since the strain induced by the uniaxial stress ($\epsilon_{aa}$, $\epsilon_{bb}$ and $\epsilon_{cc}$) does not break the D$_{2h}$ point group symmetry of ZrTe$_5$, $m$ depends linearly on strain when the strain is small. This is also consistent with DFT calculations, in which the bandgap can be approximated by $m = m_0 + 2.7 eV \epsilon_{aa} - 2.4 eV \epsilon_{cc}$. Although in the electron Hamiltonian the sign of $m$ is well defined, the resistivity does not depend on the sign. Thus, the resistivity should be an even function of the bandgap. Therefore, even for a purely linear bandgap-strain dependence near the bandgap closing, the resistivity-strain dependence will be quadratic. This is consistent with the observed symmetric quadratic elastoresistivity.

At low temperature in the quantum degenerate regime, the carrier number is practically independent of strain and determined by the number of impurities. The modulation of the bandgap by strain mainly changes the carrier mobility, which is determined by the effective mass and relaxation time of carriers at the Fermi energy. Therefore, the quadratic coefficient of the dimensionless quantity $\Delta \rho / \rho$ is determined by the bandgap $m$ normalized by the Fermi energy $E_F$:

$$\frac{\Delta \rho(\epsilon)}{\rho} \approx \left(\frac{m}{E_F}\right)^2 = \left(\frac{\frac{\partial m}{\partial \epsilon}}{E_F}\right)^2 \epsilon^2$$

A more detailed analysis based on a relativistic electron gas scattered from charged impurities leads to the same conclusion (*27*). We also measured the strain dependence of transverse magnetoresistance ($I \perp B \parallel \boldsymbol{b}$), as shown in Fig. S5. At the weak field limit the transverse magnetoresistance $\Delta \rho / \rho \sim (\omega_c \tau)^2$ peaks at $\epsilon_{aa} = \epsilon_{min}$, consistent with the description above. Using the linear coefficient of bandgap-strain dependence calculated by DFT as an input to the above formula, a fit of the quadratic coefficient yields a chemical potential of 4-8 meV and a carrier density of 0.3-2.4 × 10$^{15}$ cm$^{-3}$. These values are comparable with ARPES and transport measurements of similar samples, although our estimated carrier density is slightly lower than other reports (*22, 37*).

The validity of the above description can also be examined by the temperature dependence of elastoresistivity. The strain dependence of resistivity was measured up to 100K, as shown in Fig. 3A. Above 100K, to reach $\epsilon_{min}$ required large compressive strains such that the crystals would be bent, as evidenced by large hysteresis in the resistivity vs strain curves. Fig. 3B shows the quadratic coefficient of the parabolic fit to the resistivity-strain curves as a function of temperature for each sample. The quadratic coefficient shows a strong non-monotonic temperature dependence, with a local minimum then maximum as temperature increases. We can successfully reproduce this non-monotonic temperature dependence by a simple Boltzmann transport model that has only one free parameter – the residual carrier concentration, which is also the concentration of charged impurities [25]. The model takes into account finite temperature effect on the Fermi-Dirac distribution but considers only scattering from charged impurities. The calculated curves in Fig. 3C show excellent qualitative agreement with the measured behavior, which can be understood as the crossover between quantum degenerate and nondegenerate regimes. At intermediate temperatures $k_B T \approx E_F$, when the number of thermally excited carriers becomes comparable to the number of extrinsic carriers, the quadratic coefficient increases due to strain modulation in the number of thermally activated carriers. At even higher temperatures $k_B T \gg E_F$, thermal energy becomes the only dominant energy scale. In this regime, we expect $\Delta\rho/\rho \approx \left(E_g/k_B T\right)^2$, in agreement with the observed decrease of sensitivity as temperature increases.

In conclusion, our extensive transport measurements and detailed data analysis have revealed an exceptionally delicate topological ground state of ZrTe$_5$. An *in-situ* strain-tuned topological phase transition in this material can be easily achieved by uniaxial stress. In comparison to hydrostatic pressure, the anisotropic strain offers not only a new degree of freedom, but also a more suitable means for photoemission and scanning tunneling spectroscopy measurements. These intriguing future studies will be crucial to comprehensive understanding and precise control of topological phase transition.

**Acknowledgments:** The authors thank A. V. Balatsky, D. H. Cobden, L. Fidkowski, J. Maciejko and X. Xu, for helpful discussion. This work was supported by NSF MRSEC at UW (DMR-1719797) as well as the Gordon and Betty Moore Foundation's EPiQS Initiative, Grant GBMF6759 to J.-H.C. A. A. was also supported in part by the U.S. Department of Energy Office of Science, Basic Energy Sciences under Award. No. DE-FG02-07ER46452. W.-C.C. is supported by the Blazer Graduate Research Fellowship from the University of Alabama at Birmingham (UAB). The calculations were performed on UAB's Cheaha supercomputer. J.-H.C. also acknowledges the support from the State of Washington funded Clean Energy Institute.

Supplementary Materials:

Materials and Methods

Supplementary Text

Figures S1-S8

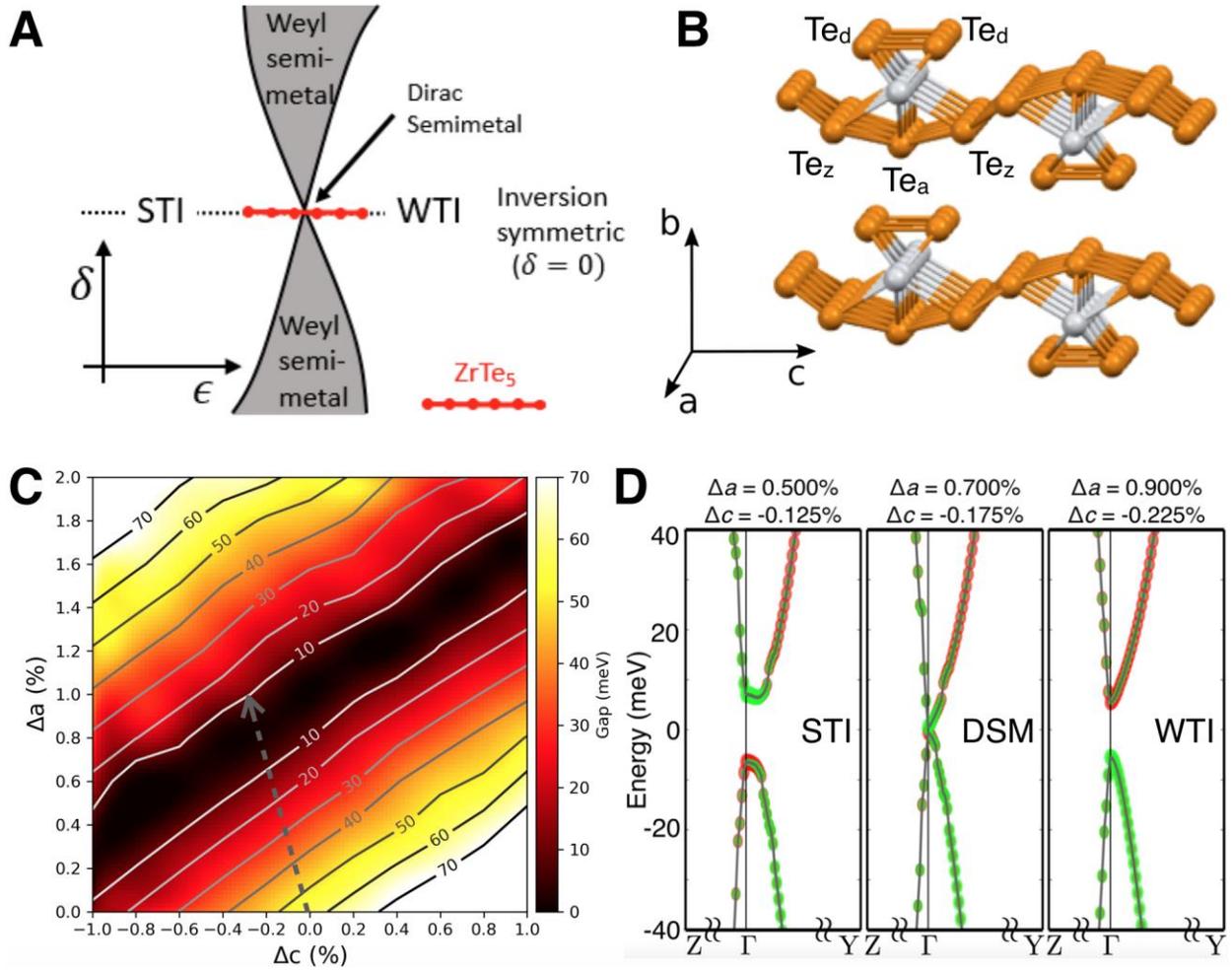

**Fig. 1 Topological phase diagram and band structures of ZrTe$_5$**. **A:** Universal Phase diagram of topological insulators proposed by Murakami for a 3D system (*3*). The control parameter $\delta$ describes the breaking of inversion symmetry. The control parameter $\epsilon$ does not break inversion symmetry. **B**: Crystal Structure of ZrTe$_5$. Chains of ZrTe$_3$ prisms (consisting of Te$_a$ and Te$_d$ atoms) extend along the *a*-axis. These chains are connected by Te$_z$ atoms along the *c*-axis to form layers. These layers are van der Waals bonded in the *b*-axis direction. **C:** The size of bandgap $E_g$ at the Γ point as functions of strains in the *a* and *c* lattice directions. The dashed grey arrow indicates the anisotropic strain induced by a uniaxial stress along the *a*-axis direction, as governed by the calculated Poisson's ratio $\epsilon_{aa} = -4.0\epsilon_{cc}$. **D** Band structures for different strain states taken at points along the Poisson's ratio path. These points (from left to right) correspond to strong topological insulator, Dirac semimetal, and weak topological insulator, respectively. Fermi level is defined as the zero energy, and the k-point labeling is based on the primitive unit cell (*12,13*). A band inversion involving Te$_d$ and Te$_z$ *p* orbitals (shown respectively as red and green colors) is seen in the strong topological insulator phase (*14*).

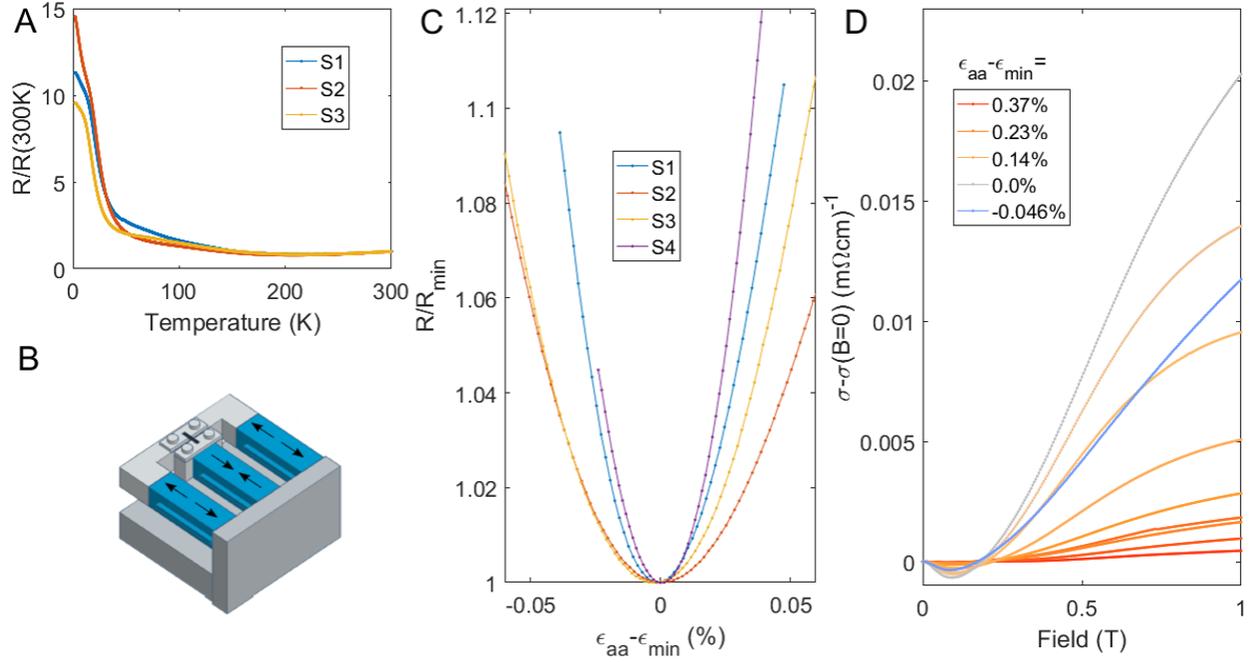

**Fig. 2: Temperature and strain dependence of transport properties of ZrTe$_5$ at T = 2K. A**: Resistivity versus temperature for three ZrTe$_5$ crystals S1-S3, as measured prior to gluing onto the 3-piezo strain apparatus. **B:** 3-piezostack apparatus used to deliver strain. **C:** Strain dependence of resistivity of ZrTe$_5$ at T = 2K for four samples S1-S4. A clear minimum in resistivity can be seen for each sample; the resistivity is normalized by its minimum value. This resistivity minimum $\rho_{min}$ varies between 1 and 16 $m\Omega cm$. The x axis is the strain along the *a*-lattice direction, which is estimated based on the method described in main text. (*27*). **D:** Positive longitudinal magnetoconductance of ZrTe$_5$ at T = 2K at different strain setpoints relative to $\epsilon_{min}$ for S3. The positive magnetoconductance peaks at the same strain value as where the resistance minimum is observed.

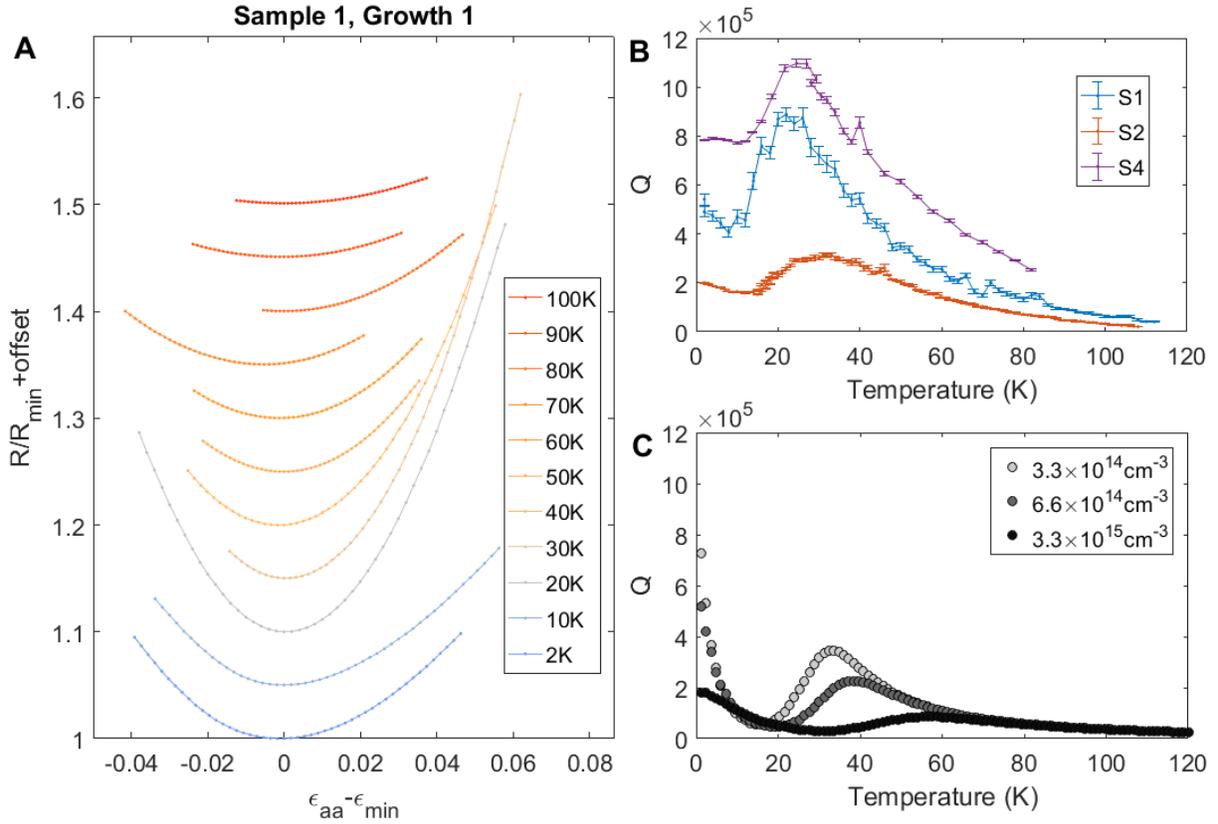

**Fig. 3 Temperature dependence of elastoresistivity. A:** Resistivity versus strain for temperatures between 2K and 100K. A clear minimum can be seen for the entire temperature range. The sensitivity of the response to strain shows a non-monotonic temperature dependence, as discussed in the main text. **B:** The resistivity-strain data fitted with a quadratic coefficient Q. The coefficient Q is plotted as a function of temperature for each crystal measured, with a local minimum then maximum as temperature is increased. **C:** The coefficient Q computed using Boltzmann transport equations (*27*). The main calculated features agree with the experimental data: a local minimum then maximum is seen with increasing temperature.

# Supplementary Materials:

# Evidence for a strain tuned topological phase transition in ZrTe$_5$

**Authors:** Joshua Mutch[1], Wei-Chih Chen[2], Preston Went[1], Tiema Qian[1], Ilham Wilson[1], Anton Andreev[1], Cheng-Chien Chen[2,*], Jiun-Haw Chu[1,*]

**Affiliations:**

[1]Department of Physics, University of Washington, Seattle, Washington 98195, USA

[2]Department of Physics, University of Alabama at Birmingham, Birmingham, Alabama 35294, USA

*Correspondence to:  jhchu@uw.edu (JHC), chencc@uab.edu (CCC)

**This PDF file includes:**

    Materials and Methods

    Supplementary Text

    Figs. S1 to S8

    Table S1

**Materials and Methods**

Material growth and sample preparation:

Single crystal $ZrTe_5$ was grown with the flux method(*20*). Zr slugs (99.9% pure, Alfa Aesar) and Te shot (99.9999% pure, Alfa Aesar) were loaded into a quartz ampule in a Zr:Te ratio of 1:100. The ampule was warmed to 900°C in 9 hours, kept at 900°C for 72 hours, then cooled to 505°C in 48 hours. To promote large crystal growth, the ampule was repetitively cooled to 440°C and warmed to 505°C. Finally, the ampule was cooled to 460°C, and decanted in a centrifuge at this temperature.

For electrical transport measurements, single crystals of typical dimensions 1.5 x 0.1 x 0.02 mm were sputtered with gold, and then 25 micron diameter gold wires were placed on the crystals and adhered with silver paint. The resistance was measured with an SRS830 lock-in amplifier and an SRS CS580 current source. Given the needle-like nature of the single crystals, the resistance was measured along the *a*-axis of the crystal.

Strain Apparatus

Uniaxial stress was applied to single crystals using a home-built 3-piezostack device, shown in Fig. S1. Three piezoelectric actuators are aligned in parallel with each other. A U-shaped Ti block was glued to the outer two piezoelectric actuators, and a small Ti block was glued to the middle actuator, forming a small gap between these blocks. Applying a voltage to the outer piezostacks while applying an equal and opposite voltage to the middle piezostack will strain the piezostacks and change the gap size.

A crystal is glued across this apparatus gap. Tuning this gap with the piezostack voltage will apply uniaxial stress to the crystal. For a similar apparatus, Hick's et al. showed that gluing only the bottom surface of the crystal to these plates can lead to strain gradients between the top and bottom surfaces of the crystal (*30*). These gradients can be suppressed by submerging the crystal in glue, as we did. Hick's et al. showed that the strain gradients are small when the ratio of $t/L_G$ (sample thickness to gap size) is small. For our measurements, this ratio is small, ranging from 0.02-0.08. We simulated the strain distribution with finite element analysis. Our finite element analysis does show that there are still some small strain gradients along the vertical axis of the crystal, mostly confined to the bottom quarter of the crystal. The resistance-strain dependence identified in this work is a smoothly-varying function. Because of this, these small strain gradients have minimal impact on our interpretation of the spatially-average resistance.

Care was taken during the construction of the 3-piezo apparatus to ensure fine alignment of the piezostacks, minimizing any stress in the secondary axes. First, a "scaffolding" piece was machined with indents the exact dimensions of the piezoelectric actuators and the Ti blocks. The actuators and blocks were placed in these indents, then glued together while secured in precise alignment. The scaffolding block was then removed after the glue dried. Second, the middle Ti block and the outer Ti block were machined with a thin flexor plate connecting them. This flexor plate restricts motion between the blocks in any axis except the primary strain axis.

Strain was measured by a foil strain gauge glued to one of the piezostacks, measuring $\epsilon^{piezo}$. The displacement strain of the device was estimated as this strain multiplied by the mechanical advantage of the apparatus, $\epsilon_{xx}^{disp} = 2L_p/L_G \epsilon^{piezo}$, where $L_P$ is the length of the piezostack, $L_G$ is the length of the gap the sample is glued across, and $\epsilon_{xx}^{disp}$ is the displacement of the apparatus. Because of a low signal to noise ratio associated with the strain gauge measurement, we present data plotted against piezostack voltage rather than plotted directly against the measured strain. We then calibrate this with a strain per volt calibration fitted from the strain measurement. We carefully minimized the amplitude of the voltage sweeps applied to the piezostacks, staying centered around $\epsilon_{min}$ to minimize any hysteresis effect in the piezostacks.

Finite Element Analysis

We used the *ANSYS Academic Research Mechanical 19.1* finite element analysis software to calculate the strain transmission in the primary axis and deformation in the secondary axis for all of our ZrTe$_5$ samples. The model is shown in Fig. S2. The stiffness tensor for ZrTe$_5$ was taken from a previously calculated value by the Materials Project (*38*). The sample is modelled as mounted to the strain apparatus in a puddle of epoxy. The Young's Modulus and Poisson's ratio of the epoxy are given the same values as the work done by Hicks (*30*). For each crystal measured, the model took the exact crystal dimensions and gap size of the apparatus as inputs, and computed an average relaxation constant $\alpha$, defined as $\epsilon_{aa} = \alpha \epsilon_{xx}^{disp}$ for $\epsilon_{xx}^{disp} = \pm 0.1\%$, where $\epsilon_{aa}$ is the strain delivered to the crystal. Our results are shown in Table S1.

Experimental Procedure

After cooling to 2K, the sample was warmed by incremental temperature set points. In-situ stress was applied to the crystal at each temperature set point by applying a triangle voltage waveform across the outer two piezo actuators, and an equal and opposite sign waveform across the middle actuator. For each temperature set point, the voltage waveform was allowed to loop several times to inspect any hysteresis effects. The magnitude and offset of this waveform was adjusted during warming to stay centered around the resistance minimum. At temperatures 80K and higher, it took an increasing amount of compressive strain to stay centered on the resistance minimum. At these temperatures the samples often buckled as negative strain was applied. This buckling led to a large hysteresis developing in the resistance-strain relation, and the experiment was terminated.

Calibration to Zero-Strain State

One of the advantages of the 3-piezo apparatus compared to direct gluing to a single piezostack is that thermal strain is minimized. This is because the large thermal expansion of the outer two piezostacks is compensated by the expansion of the middle piezostack. However, there is still a non-negligible thermal strain resulting from the mismatch of thermal expansions between the crystal and the titanium pieces of the apparatus. Since titanium is known to have a smaller thermal expansion compared to most materials, it is expected that cooling the apparatus will impart a tensile strain to a mounted crystal. By tuning the controllable strain of the apparatus, this thermal strain can be compensated if a reference calibration is available.

A zero-strain calibration was constructed by measuring the resistance of crystals prior to gluing to the strain apparatus. After measuring the zero-strain resistance, crystals were glued to the apparatus and the resistance was measured while cooling from 300K to 2K. The resistance of a crystal mounted on the apparatus was plotted real-time during cooling, overlaid against the zero-strain calibration. As thermal strain became significant, the resistance of the strained crystal deviated from the zero-strain calibration resistance. The apparatus strain could then be adjusted to tune the strained resistance to the calibration resistance, keeping the crystal in the zero-strain state. We were always able to track the zero-strain state of the crystal down to about 60K. Between 60K and 300K, crystals always had a positive gauge factor (tensile strain increased resistance). This indicates that in this temperature range the zero-strain state resides at higher strains compared to $\epsilon_{min}$.

Below 60K, two things occurred that made tracking zero-strain state difficult. The first was that the resistance sensitivity to strain became weak - the gauge factor approached zero. This indicates that the crystal either approaches or passes through $\epsilon_{min}$ below 60K. The second difficulty is that the resistance of the mounted crystal measured a slightly higher value than the calibration resistance for temperatures below 60K. This occurred even when the strained crystal was tuned to $\epsilon_{min}$.

The reason for the inconsistency between strained resistance and the zero-strain calibration resistance below 60K is unknown. There might be an effect due to the residual strain induced by the thermal contraction of the epoxy used to glue the samples.

We were able to rule out two possible explanations for this inconsistency: stray magnetic fields and thermal lag. Since the magnetoresistance of ZrTe$_5$ is large, it was considered that stray magnetic fields of order 40-100 Oe might explain this discrepancy. The transport measurements were taken in a Quantum Design Dynacool 14T system, which is known to have stray magnetic fields on the order of 40-100Oe resulting from flux-pinning if the magnetic field is linearly ramped to zero. By oscillating the field to zero from 2T, these stray fields can be greatly suppressed. To eliminate stray fields as a possible source of this inconsistency, calibration and strain measurements were performed after the field was oscillated to zero. This inconsistency persisted even after these precautions. Thermal lag as a source of this inconsistency was also eliminated by measuring resistance at setpoints rather than a slow temperature ramp. This, too, did not resolve the inconsistency.

To estimate the location of the zero-strain state, i.e. the absolute value of $\epsilon_{min}$, we performed the following experiment. Crystals were glued directly on the side wall of a single piezostack, as shown in Fig. S4B and S4C. The piezostack has unusual highly anisotropic thermal expansion properties; it expands by about 0.1% along the polling direction and contracts along the transverse direction as it is cooled to liquid helium temperatures. Gluing crystals oriented parallel and perpendicular to the piezostack polling axis imparts a very different strain during cooling, mimicking scenarios where samples were glued on substrates with different thermal expansion coefficients. Tuning the piezostack voltage at 2K adds a much smaller tunable strain (~ 0.01 – 0.02%) on top of this thermal strain. Using this tunable strain we are able to measure a linear elastoresistance. The slope of this linear response, defined as the gauge factor $GF = (\frac{\Delta\rho}{\rho})/(\frac{\Delta L}{L})$, measures the local derivative of the nonlinear resistivity vs strain curves. A positive or negative gauge factor indicates which side of $\epsilon_{min}$ the thermally strained crystal resides on. As seen in Fig. S4, the parallel (perpendicular) orientations measure a positive (negative) $GF$ at 2K. This indicates the sensitivity of the thermal strain to sample preparation, and allows us to make an estimate of $\epsilon_{min}$. Based on published data for the thermal expansion of similar piezostacks and ZrTe$_5$, cooling to 2K strains the perpendicular glued sample by about +0.08% (*25, 39*). The parallel glued sample is strained even more than this. We measured $GF(2K) = -73$ for the perpendicular glued sample. This indicates that the parallel glued thermal strain is between $-0.04\%$ to $-0.01\%$ with respect to $\epsilon_{min}$, as calibrated by the quadratic response we measured for samples in this work. Combining this with our estimate for the thermal strain, we estimate $\epsilon_{min}$ is at most +0.12% at 2K.

Transverse Magnetoresistance

Magnetoresistance (MR) was measured at strain setpoints for field parallel the *b*-axis and current along the *a*-axis. A quantity with the dimensions of mobility can be obtained by fitting $\frac{\rho(B)}{\rho_0} = 1 + \mu^2 B^2$ to magnetoresistance data. This expression for the mobility is exact in a two-band model in the limit of equal electron and hole charge carrier densities with the same mobilities. The MR of ZrTe$_5$ for magnetic fields aligned in this manner has been reported before, and does not fit a quadratic dependence for low temperatures and high fields (*40*). However, we found that for extremely small fields (<75 Oe), the MR can be fit to a quadratic dependence at 2K, as shown in Fig. S5. The mobility obtained from the MR data is shown to be a non-monotonic function of strain, with a maximum near $\epsilon_{min}$. At low temperatures, when $k_B T \ll E_F$, this maximum in mobility near $\epsilon_{min}$ causes the measured minimum in resistance at the bandgap closing point rather than thermally excited carriers.

The low-field data shown in Fig. S5 was collected using a Quantum Design Dynacool 14T system. Prior to measurements the magnetic field was "oscillated" to zero from 2T to suppress flux pinning in the Dynacool magnet, giving a more accurate measurement of the magnetic field.

Longitudinal Magnetoresistance

Longitudinal MR was measured at strain setpoints for field parallel the current along the *a* axis. In addition to the results reported in Fig. 2D in the main text, another crystal was measured for strains very close to $\epsilon_{min}$, shown in Fig. S6.. The positive magnetoconductance was suppressed for strains as low as 0.02% away from $\epsilon_{min}$, for both compressive and tensile strains.

Boltzmann transport theory of massive Dirac semimetal

At low temperature scattering is dominated by charged impurity scattering, and the mean free path $l$ is determined by the impurity concentration $n_i$ and the impurity trans cross-section $\sigma_{tran}$:

$$\frac{1}{l} = n_i \sigma_{tran}$$

Near the Γ point, the band structure of ZrTe$_5$ obeys a relativistic Dirac dispersion. The scattering of this relativistic dispersion by a Coulombic potential obeys the relativistic Mott formula:

$$\sigma_{tran} = \int do \frac{d\sigma}{do}[1 - \cos\theta]$$

$$\sigma_{tran} = \frac{\pi\alpha^2}{k^2}\int_{-1}^{1} d\cos\theta \frac{\beta^{-2} - \sin^2\theta/2}{\sin^2\theta/2}$$

$$\sigma_{tran} \sim \frac{8\pi\alpha^2}{k^2\beta^2}\ln\frac{1}{\theta_0}$$

Where $\sigma_{tran}$ is the scattering cross section, $\theta_0$ is the small angle cutoff due to the screening of the Coulomb interaction, and $\alpha = \frac{e^2}{\hbar v \epsilon}$ is the dimensionless coupling constant, where $\epsilon$ is the static dielectric constant. $\beta = v^2 p^2/(m^2 + v^2 p^2)$ follows from $\beta = \frac{1}{\hbar v}\frac{\partial E}{\partial k}$ applied to the dispersion $E^2 = m^2 + \hbar^2 k^2 v^2$. At low temperatures, the dominant effect of bandgap size on resistivity is the change in scattering cross section. Since $\rho \propto \sigma_{tran}(m)$, it follows that

$$\frac{\rho(m) - \rho(m=0)}{\rho(m=0)} \approx \frac{\beta(m=0) - \beta(m)}{\beta(m)} = 1 + \left(\frac{m}{E_F}\right)^2$$

Which is the result in the main text. The carrier density for each sample measured based on this computation is estimated in Table S1.

Temperature Dependence of Resistivity/Gap Relation

A semiclassical Boltzmann transport model is used to compute the temperature dependence of the resistivity sensitivity to strain. The Dirac band at the Γ point is assumed to dominate conduction, and the conductivity can be expressed as:

$$\sigma_{xx} = -e^2 \int g(E)\tau(E)v_x(E)^2 \frac{\delta f}{\delta E} dE$$

Where $v_x = \frac{1}{\hbar}\frac{\partial E}{\partial k_x}$. The scattering time $\tau$ can be expressed as $\frac{l(E)}{v_x}$, where $l(E)$ is the energy dependent scattering length mentioned above, inversely proportional to $\sigma_{tran}$. We assume that scattering off of charged impurities is the dominant scattering mechanism in this simplistic model. Given the Dirac dispersion of $E^2 = m^2 + \hbar^2 k^2 v^2$, the density of states can be written as

$$g(E) = \frac{E\sqrt{E^2 - m^2}}{\pi^2 \hbar^3 v^3}.$$

Combining these, the conductivity may be written as

$$\frac{\sigma_{xx}(m)}{\sigma_{xx}(m=0)} = \left( \int \frac{(E^2 - m^2)^3}{E^2} \frac{1}{2\cosh\left(\frac{E-\mu}{k_B T}\right) + 2} dE \right) \div \left( \int \frac{E^4}{2\cosh\left(\frac{E-\mu}{k_B T}\right) + 2} dE \right)$$

Given the chemical potential $\mu$ as a function of temperature, $\frac{\sigma_{xx}(T,m)}{\sigma_{xx}(T,m=0)}$ can be computed for a range of temperatures, and a quadratic relation between $1/\sigma_{xx}$ and $m$ can be fitted.

All that is left in the model is to compute $\mu(T)$. This can easily be done by charge conservation. The impurity doping $n_i$ is defined as $n - p$, where $n$ and $p$ are the electron and hole densities. $n(T)$ and $p(T)$ can be computed as:

$$n = \int_0^\infty g(E) \frac{1}{e^{E-\mu} + 1} dE$$

$$p = \int_{-\infty}^0 g(E) \left(1 - \frac{1}{e^{E-\mu} + 1}\right) dE$$

Given $n_i$ as the input parameter, $\mu(T)$ can be solved numerically by enforcing $n_i = n(T) - p(T)$ for electron doping or $n_i = p(T) - n(T)$ for hole doping. The only other input parameter for our computational model is the velocity $v$ that appears in the Dirac dispersion. We used $v_a, v_b, v_c = 1.7, 5.2, 2.2 \times 10^5 m/s$, as measured by SdH oscillations (*18*).

Density Functional Theory Calculations

The electronic structure and energy gap of ZrTe$_5$ under strain are calculated by the Quantum Espresso package (*41*) based on density functional theory. The Perdew-Burke-Ernzerhof exchange-correlation functional (PBE-GGA) (*42*) with spin-orbit coupling and projector-augmented-wave (PAW) (*43*) method are used. The unstrained lattice parameters are determined by experimental data at 10K (*25*), and an 11×11 grid in the parameter ranges of 1.0$a$ - 1.02$a$ and 0.99$c$ - 1.01$c$ is considered for studying the gap behavior with lattice variation. The raw data of the zone-center gap are shown in Fig. S7A, and the interpolated data with two-dimensional cubic splines are shown in Fig. 1C in the main text. An 8×8×4 momentum grid is used in the self-consistent calculation, and the kinetic energy cutoff and convergence criterion are set to 30 Ry and 10$^{-7}$ Ry, respectively. The DFT band structures for ZrTe$_5$ in different strained states are shown in Fig. S8. The labeling of the high-symmetry k-points is based on the Brillouin zone of the primitive unit cell.

The topological $Z_2$ indices of different strained structures are also computed to identify their topological nature. With input from the electronic structure calculations of Quantum Espresso, the maximally localized Wannier functions are first computed using the Wannier90 package (*44*), which in turn allows the determination of $Z_2$ indices by tracking hybrid Wannier charge centers using the WannierTools package (*45*). As shown in Fig. S7B, the bottom right of the phase diagram is a strong topological insulator (STI with $Z_2$ indices (1;110)), and the upper left is a weak topological insulator (WTI with $Z_2$ indices (0;110)). Phase transition between the STI and WTI states is directly controlled by closing the energy gap at the Brillouin zone center.

Additional structure relaxation calculations are performed with the van der Walls density functional theory (vdW-DFT) (*46, 47*) corrected using the exchange-hole dipole moment model (*48*). The vdW-DFT fully-relaxed lattice parameters of layer-structured $ZrTe_5$ are within 1% error compared to the unstrained experimental data. To determine the Poisson ratio, conventional cells of different fixed lattice parameters along the *a*-axis are considered, while the *b* and *c* lattice parameters are allowed to evolve freely in the structure relaxation calculations. The results are shown in Fig. S7 C.

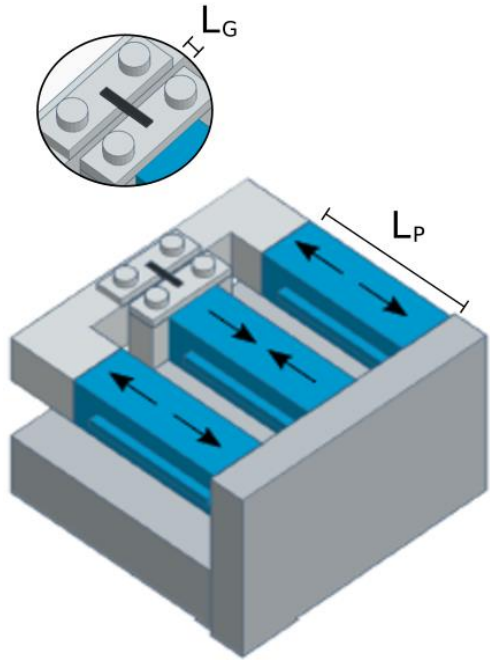 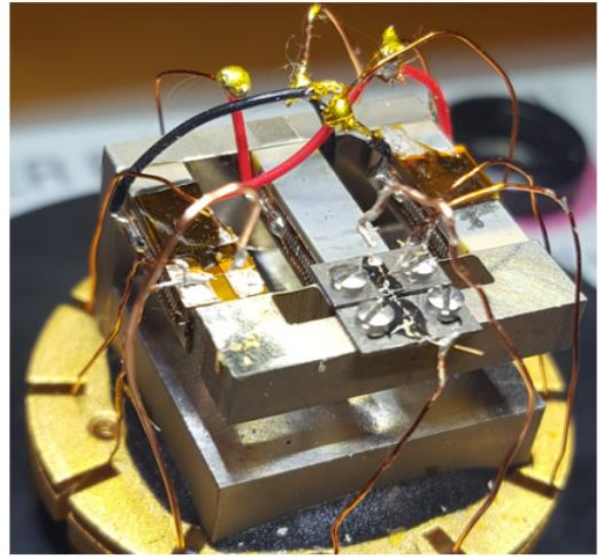

**Fig. S1:** 3-piezo device. By applying inducing strain in the outer and middle actuators (blue), stress can be applied to a crystal glued across the gap (black). The displacement of the gap is a factor of of $2L_P/L_G$ larger than the strain of any single piezostack.

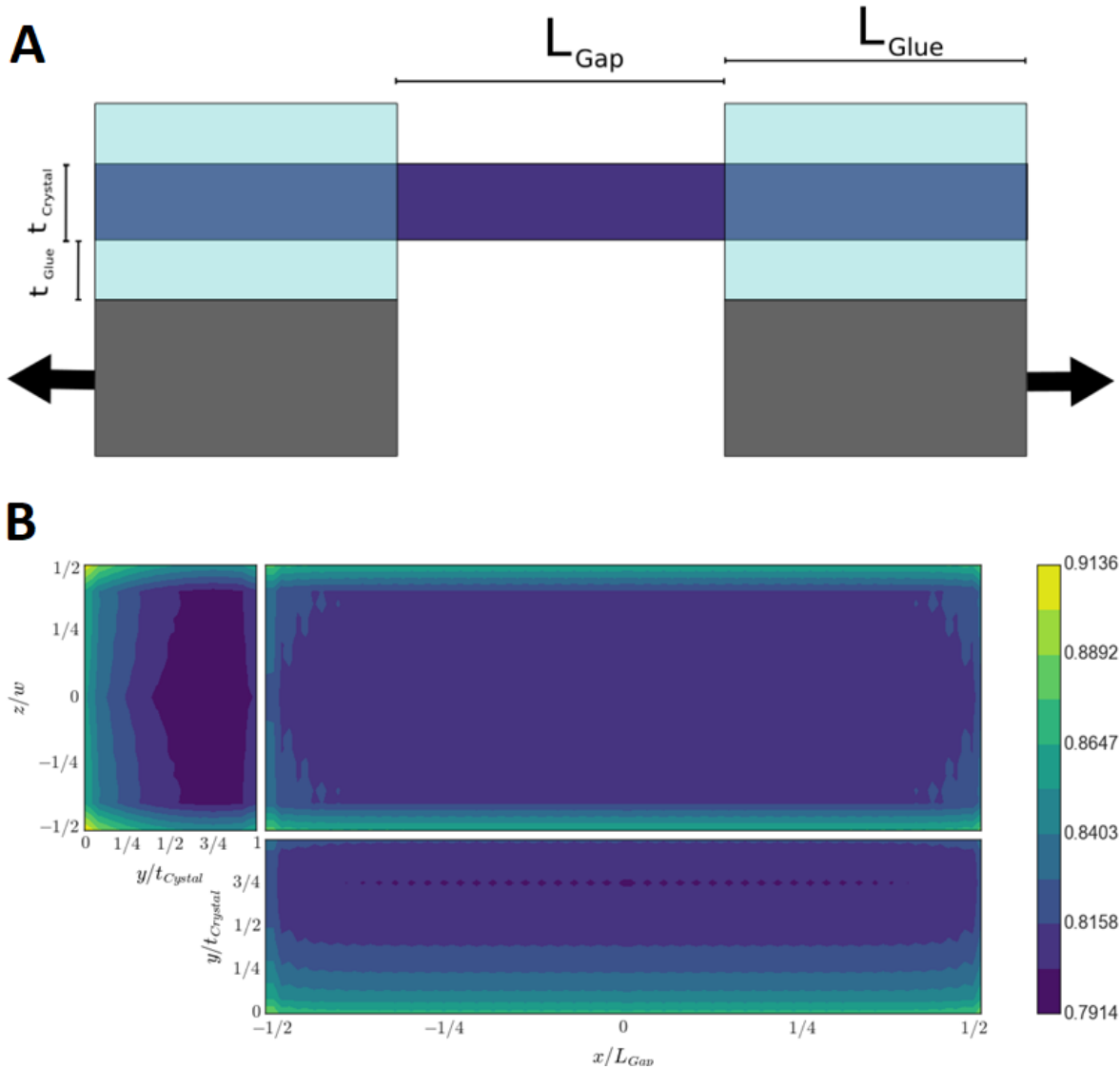

**Fig S2: A:** Schematic of system modelled by finite element analysis. For our analysis, we assumed $t_{Glue}/t_{Crystal}=0.5$ and $L_{Glue}/L_{Gap}=1$, reasonable assumptions given optical images of the experiment. **B:** Strain transmission $\alpha$ averaged through the length (upper left), thickness (upper right), and width (bottom right) of the crystal, modelled after a crystal with dimensions of Growth 1, Sample 2.

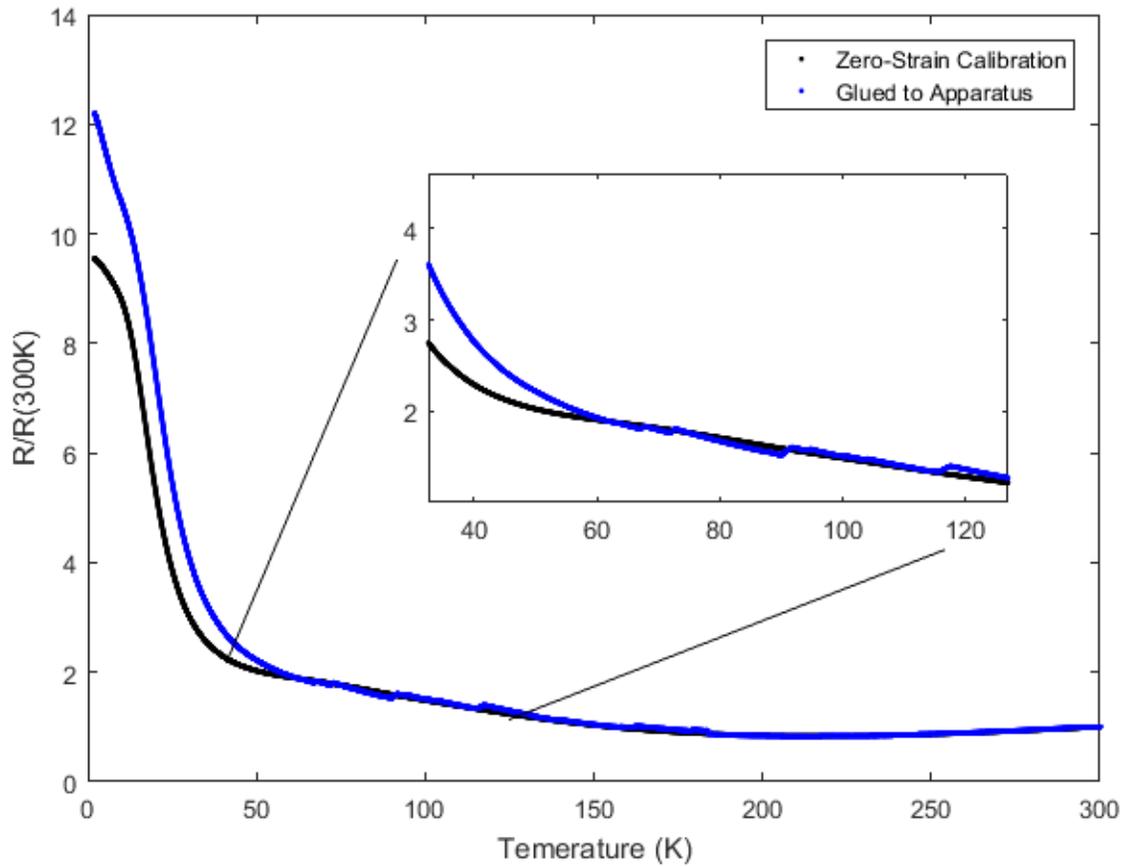

**Fig. S3:** Resistance versus temperature for a ZrTe$_5$ crystal, before (black) and after (blue) being glued to the strain apparatus. Discontinuities in the slope of the resistance indicate the apparatus strain being tuned in an attempt to minimize the thermal strain. Near 60K the calibration to the zero-strain state was lost, due to reasons discussed. Although the 2K difference between the glued and zero-strain state appears large (>20%), the resistance at 2K can be tuned to less than a 4% difference from the calibration data.

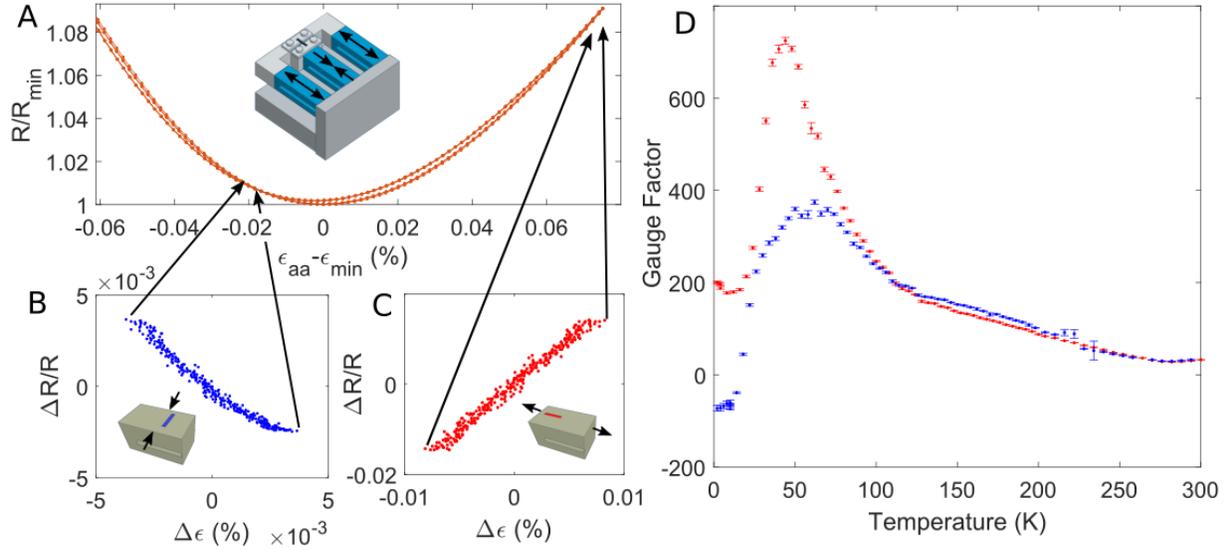

**Fig. S4 A:** Resistance versus strain at 2K for sample S2 mounted on the 3-piezo mechanism. **B & C:** Resistance versus strain at 2K for samples glued directly to the surface of a piezostack, glued perpendicular (B) and parallel (C) to the polling direction of the piezostack. **D:** Gauge factor, defined as the linear slope of resistance as a function of strain, for the two crystals glued perpendicular and parallel the piezo polling direction. Below 100K, the GF becomes sensitive to the sample mounting method.

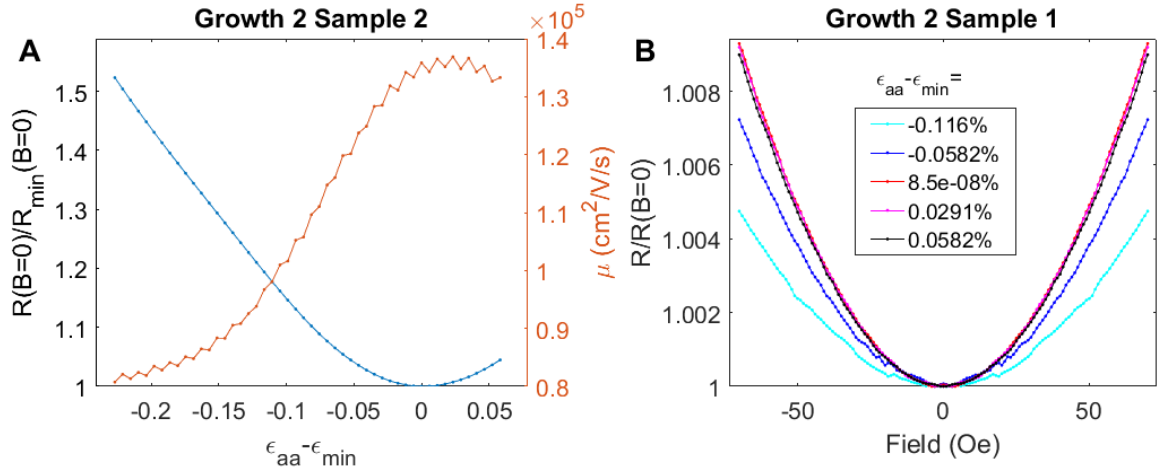

**Fig. S5 A:** Resistance (left axis) and mobility (right axis) as a function of strain. **B:** Magnetoresistance data used for the fitting of $\rho(B)/\rho_0 = 1 + \mu^2 B^2$. A peak response in the sensitivity to the field can be seen near $\epsilon_{min}$.

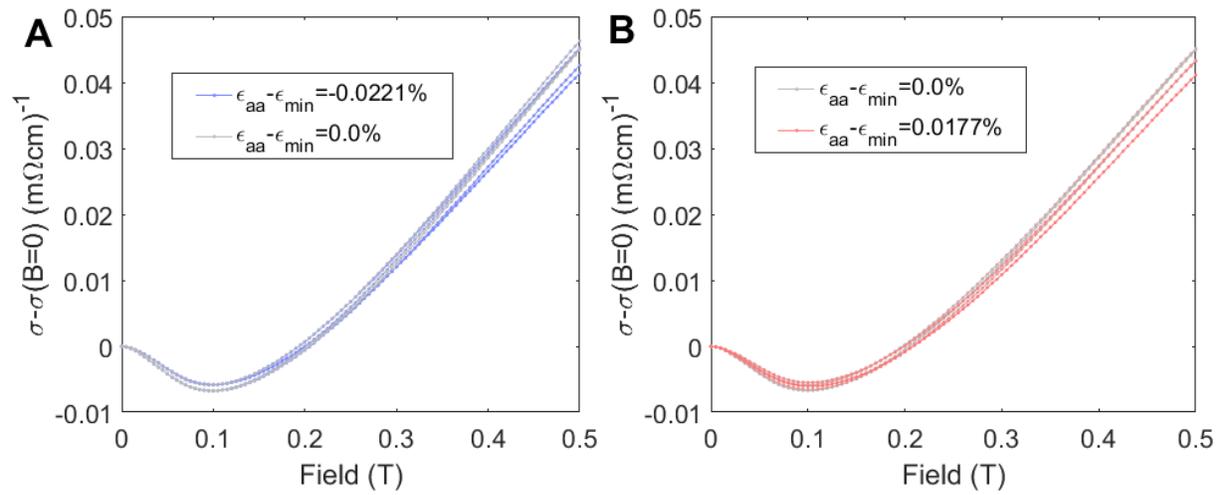

**Fig. S6** Positive magnetoconductance for compressive **(A)** and tensile **(B)** strains with respect to $\epsilon_{min}$. The strength of the positive magnetoconductance is suppressed for strains away from the $\epsilon_{min}$.

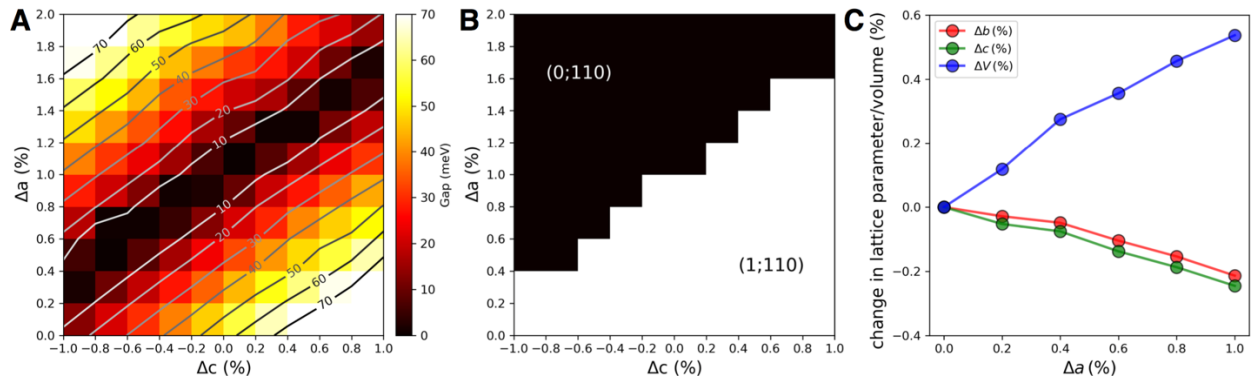

**Fig. S7 A:** Zone-center energy gap computed by DFT as functions of variations in the lattice *a* and *c* parameters. **B:** Topological phase diagram and $Z_2$ indices for different strained structures. Phase transition between the STI (1;110) and WTI (0;110) states is directly controlled by closing the zone-center energy gap. **C:** Changes in the *b/c* lattice parameters and volume as a function of strain in the *a*-direction; the fully relaxed structure calculations were performed with the van der Walls density functional theory (vdW-DFT).

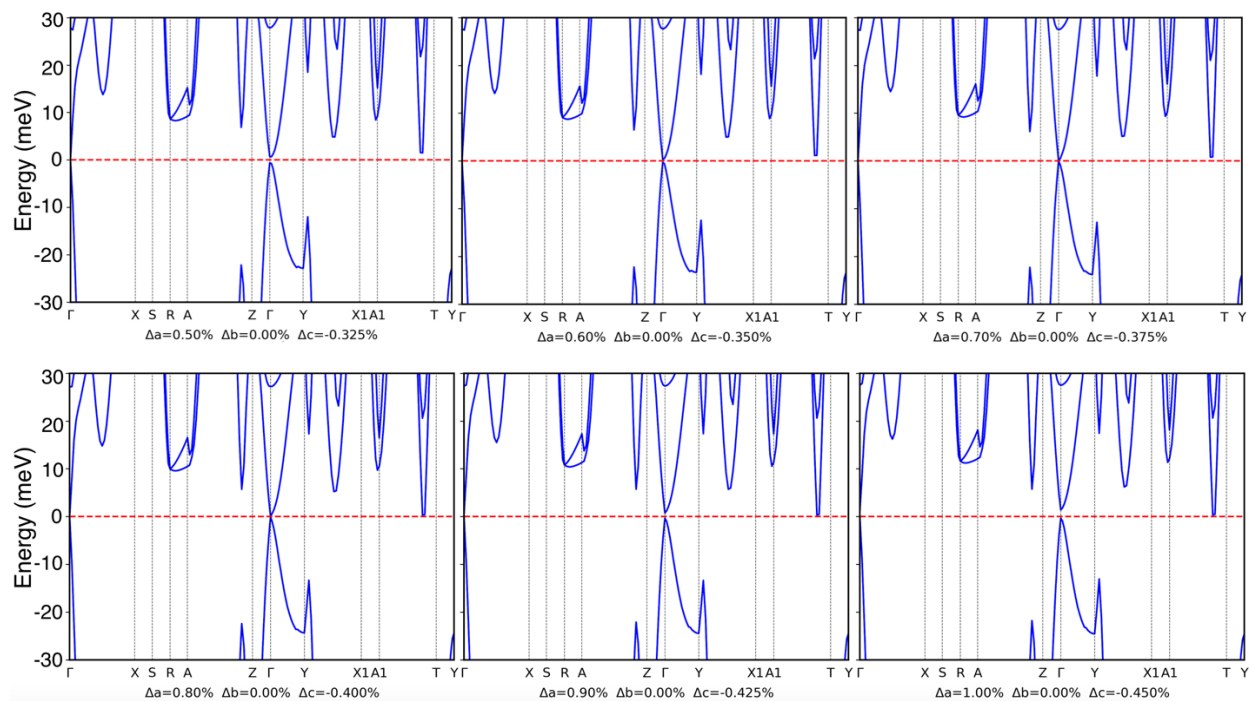

**Fig. S8** DFT band structures for ZrTe$_5$ in different strained states. The horizontal red dashed line indicates the Fermi level. The labeling of the high-symmetry k-points is based on the Brillouin zone of the primitive unit cell.

| Sample | Growth | ($L_G$:$w$:$t$) (μm) | α | $Q$(2K) | $n$ (cm$^{-3}$) |
|---|---|---|---|---|---|
| S1 | 1 | 962:60:10 | 0.93 ± 0.06 | $4.9 \times 10^5$ | 0.6 |
| S2 | 1 | 1,060:120:40 | 0.81 ± 0.08 | $2.1 \times 10^5$ | 2.4 |
| S3 | 1 | 490:21:15 | 0.89 ± 0.07 | $2.7 \times 10^5$ | 1.6 |
| S4 | 2 | 800:40:60 | 0.82 ± 0.07 | $7.8 \times 10^5$ | 0.3 |

**Table S1**: Dimensions of apparatus gap ($L_G$), crystal thickness, and crystal width are reported for each crystal in this work. The average strain relaxation $\alpha$ is calculated from these parameters by finite element analysis. The deformation in the vertical axis, $\Delta y$, is also calculated. The quadratic response to strain is denoted as $Q$. The carrier density $n$ calculated from $Q$(2K) is numerically computed from a Boltzmann transport equation of the Dirac dispersion.